\documentclass[prl,groupedaddress,showpacs,twocolumn]{revtex4}
\usepackage{graphicx}
\usepackage{amssymb}
\newcommand{\ped}[1]{\ensuremath{_{\rm #1}}}
\newcommand{\apex}[1]{\ensuremath{^{\rm #1}}}

\begin{document}

\title{Large conductance modulation of gold thin films by huge charge injection via electrochemical gating}

\author {D. Daghero \email{E-mail:dario.daghero@polito.it}}
\author {F. Paolucci}
\altaffiliation{Now at Max Planck Institute for Solid State
Research, Stuttgart (Germany).}
\author{A. Sola}
\author {M.~Tortello}
\author {G.A. Ummarino}
\author {R.S. Gonnelli}
\affiliation {Dipartimento di Fisica, Politecnico di Torino, 10129
Torino, Italy}
\author {Jijeesh R. Nair}
\author {C. Gerbaldi}
\affiliation {Dipartimento di Scienza dei Materiali e Ingegneria
Chimica, Politecnico di Torino, 10129 Torino, Italy}


\begin{abstract}
By using an electrochemical gating technique with a new combination
of polymer and electrolyte, we were able to inject surface charge
densities $n_{2D}$ as high as $3.5 \times 10^{15}$ e/cm$^{2}$ in
gold films and to observe large relative variations in the film
resistance, $\Delta R/R'$, up to 10\% at low temperature. $\Delta
R/R'$ is a linear function of $n_{2D}$ -- as expected within a
free-electron model -- if the film is thick enough ($\geq 25$ nm),
otherwise a tendency to saturation due to size effects is observed.
The application of this technique to 2D materials will allow
extending the field-effect experiments to a range of charge doping
where giant conductance modulations and, in some cases, even the
occurrence of superconductivity are expected.
\end{abstract} \maketitle

Since the Sixties, the possibility to modulate the transport
properties of various materials by means of the so-called field
effect (FE) has attracted much interest. Apart from the nowadays
obvious application in semiconductor-based electronic devices such
as FETs (field-effect transistors), the technique has been widely
used also for more exotic purposes. It has allowed enhancing the
critical temperature of some superconductors
\cite{Glover60,Mannhart91,Ahn99}, inducing metallic behavior in
insulators \cite{Shimotani07} or even a superconducting phase
transition in materials like SrTiO\ped{3} \cite{Ueno08} ZrNCl
\cite{Ye09} and KTaO$_3$ \cite{Ueno11}. In the standard FET
configuration, the maximum density of the induced surface charge,
$\sigma_{max}$, is of the order of $10^{13}$ charges cm$^{-2}$ if
suitable dielectrics are used. Only with a polymeric gating
technique \cite{Panzer05,Dhoot06} electric fields as high as
$100\mathrm{MV/cm}$, and surface carrier concentrations of
$10\apex{14}/\mathrm{cm}\apex{2}$ \cite{Ye09} have been achieved.
The present record, to the best of our knowledge, is
$4.5\times10^{14}$ cm$^{-2}$ \cite{Yuan09}. The reason of this
order-of-magnitude improvement with respect to the conventional FETs
is the formation of the electric double layer (EDL) at the interface
between the electrolyte solution and the sample surface. The EDL
acts as a parallel-plate capacitor with extremely small distance
between the plates (of the order of the polymer molecule size)
\cite{Ye09} and thus very large capacitance.

Here, we will show that a new polymeric electrolyte solution (PES)
allows further extending the surface charge density to some units in
10$^{15}$ charges cm$^{-2}$, for applied voltages of the order of a
few Volts (5 V at most), which marks a significant improvement with
respect to the present state of the art. In particular, we will
apply this technique to Au films.

The FE in metals has been devoted little attention, either because
of its little practical interest or because often believed to be
unobservable. Indeed, in the semiclassical, metallic limit, the
electronic screening length (the Thomas-Fermi radius) is less than
one atomic diameter. Nonetheless, a modulation of the conductivity
of metal films (including Au) has been obtained already in the
Sixties \cite{Bonfiglioli59, Bonfiglioli65} with a conventional
gating technique. These and the following measurements of the same
kind \cite{Glover60,Stadler65,Martinez97,Markovic01} have evidenced
a number of unexpected properties and differences between metals
that well justify a fundamental interest in this topic -- especially
because most of these results have not found a really exhaustive
explanation up to now.

We will leave the fundamental study of the FE in gold and other
metals (Cu, Ag) to a following paper. Here we will just focus on the
technique that allows extending the field-effect studies to
unprecedented surface charge densities. In particular, we will show
that this technique allows observing very large modulations in the
gold resistivity both at room temperature and at cryogenic
temperatures. The relative variation of the film resistance $\Delta
R/R'$ produced by the transverse electric field can be as high as
10\% at low temperature and perfectly extends the analogous results
obtained at much smaller charge densities by using the standard FET
configuration.

\begin{figure}[t]
\begin{center}
\includegraphics[keepaspectratio, width=0.8\columnwidth]{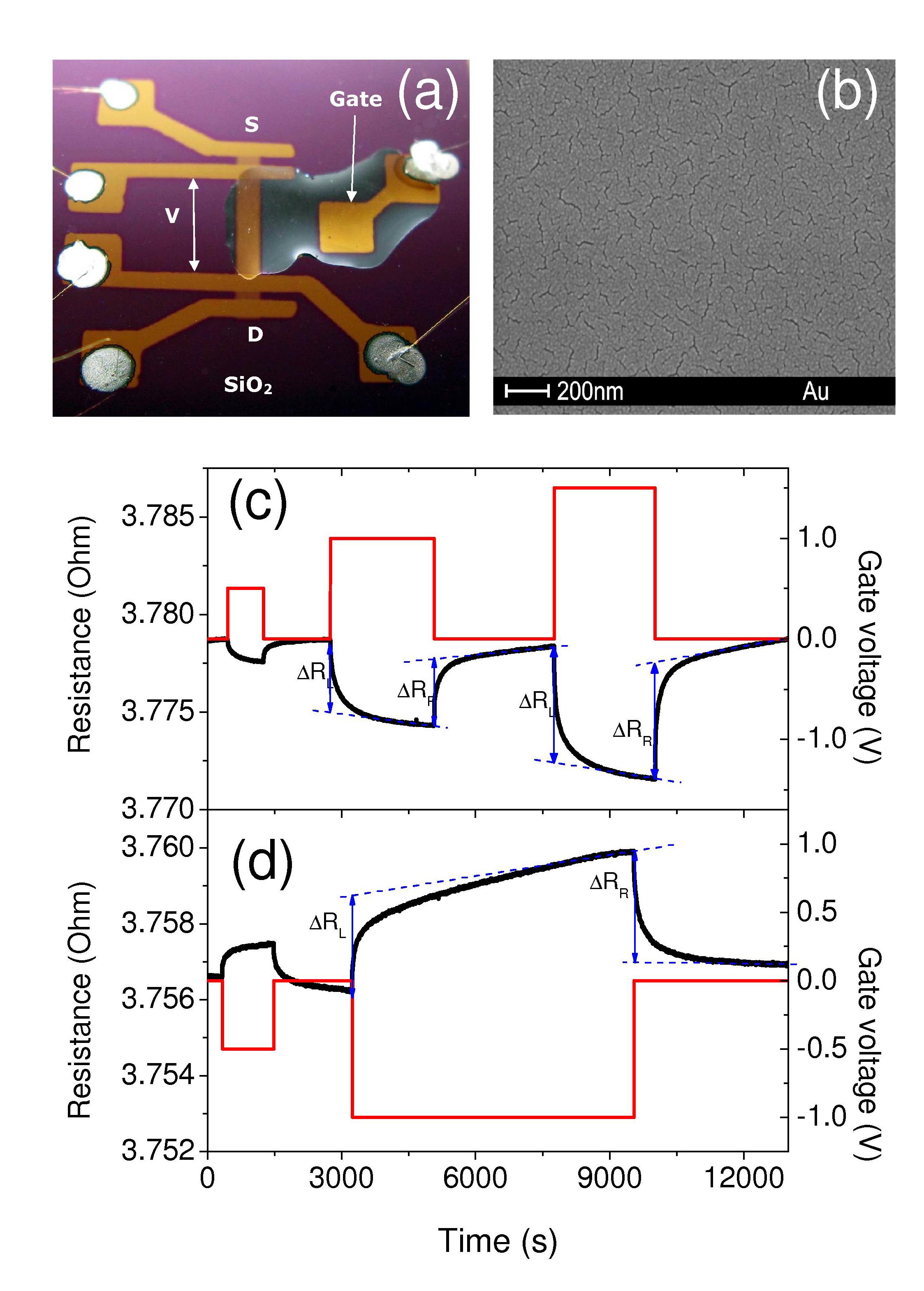}
\end{center}
\vspace{-7mm}\caption{(a) Photograph of a Au FED on SiO$_2$
substrate. D and S are the drain and source contacts; the voltage is
measured between the inner contacts. The drop of polymer electrolyte
covers the part of the film between the voltage contacts as well as
the gate electrode. (b) SEM image of the Au film. (c,d) Typical
response of the film resistance to positive and negative gate
voltages.} \label{fig:1}
\end{figure}

The field-effect devices (FEDs) were fabricated on glass, SiO$_2$ or
Si$_3$N$_4$ substrates and were designed in a completely planar
configuration, as in ref. \cite{Dhoot06}, with the film under study
and all the electrodes (drain, source, contacts for voltage
measurement and gate) on the same plane. A picture of a device on
SiO$_2$ is shown in fig. \ref{fig:1}a.

The gold films were deposited by physical vapour deposition (PVD) at
a pressure $P \sim 2 \cdot 10\apex{-5}$ mbar, in the forms of a thin
strip. The thickness of the films, measured by means of a
profilometer and/or an atomic force microscope (AFM), ranges between
10 and 50 nm. SEM images of the film surface (fig.\ref{fig:1}b) show
accretion islands connected to form a continuous network. This kind
of structure is typical of the best gold films grown by PVD, as
reported in literature \cite{Zhang99}. The four gold electrodes for
current feeding and voltage measurement, as well as the gate
electrode, were then deposited on top of the film by PVD at $P \sim
4 \cdot 10\apex{-5}$ mbar, and are much thicker than the film. The
polymer electrolyte solution we used was obtained by UV-curing a
reactive mixture of bisphenol A ethoxylate (15 EO/phenol)
dimethacrylate (BEMA, average M\ped{n}: 1700, Aldrich),
poly(ethylene glycol)methyl ether methacrylate (PEGMA, average
M\ped{n}: 475, Aldrich) and lithium
bis(trifluoromethanesulfonyl)imide (LiTFSI) in the presence of
$2\%$wt of 2-hydroxy-2-methyl-1-phenyl-1-propanon (Darocur1173, Ciba
Specialty Chemicals) free radical photo-initiator. The quantity of
BEMA and PEGMA are in $3:7$ ratio and the LiTFSI is the $10\%$wt of
the total compound.

The PES was put on top of the device, in such a way that the whole
portion of the film between the voltage electrodes as well as the
gate electrode were covered, as shown in Fig.\ref{fig:1}a. Since the
area of the gate electrode is larger than that of the film, there is
no need of reference electrode \cite{Yuan10}. A photochemical curing
was then performed by using a medium vapor pressure Hg UV lamp
(Helios Ital quartz, Italy), with a radiation intensity on the
surface of the sample of 28 mW cm$^{-2}$. All the above operations
were performed in controlled Ar atmosphere of a dry glove-box
(MBraun Labstar, O$_2$ and H$_2$O content $< 0.1$ ppm) .

The field-effect devices were then mounted in a pulse-tube
cryocooler and kept in high vacuum to protect the PES from moisture
and chemical contaminations. Figures \ref{fig:1}(c) and (d) show the
effect of positive and negative voltage steps (applied at
$T_{room}=295$ K, above the glassy transition of the polymer that
occurs at about 210 K) on the resistance of the film, measured with
the four-terminal technique with a DC current of 1-5 mA and by
inverting the current to eliminate thermoelectric effects. The film
resistance is related to the applied voltage through the charge on
the EDL. For a given gate voltage $V_{G}$ the resistance variation
$\Delta R=[R(V_{G})-R_0]$ (where $R_0=R(V_G=0)$) is obtained by
averaging the resistance jumps $\Delta R_R$ and $\Delta R_L$ on
applying and removing the gate voltage, as shown in fig.
\ref{fig:1}.

The problem then arises of how to relate the gate voltage to the
charge of the EDL and thus the density of the surface charge
injected in the film. Hall-effect measurements would require huge
magnetic fields because of the high intrinsic carrier density of Au.
Moreover, determining the charge of the EDL by integrating the gate
current is not correct if electrochemical effects are present, as
pointed out in ref. \cite{Yuan10}. Electrochemical Impedance
Spectroscopy (EIS) measurements carried out both on our devices and
on a steel/PES/steel cell showed indeed that electrochemical effects
take place at frequencies below 10 Hz \cite{Yuan10}. We thus used a
procedure called \emph{double-step chronocoulometry} \cite{Inzelt10}
that allows separating the electrostatic charge we are interested in
from the charge that flows through the PES because of
electrochemical effects (e.g. diffusion of electroreactants).

\begin{figure}[t]
\begin{center}
\includegraphics[keepaspectratio, width=0.7\columnwidth]{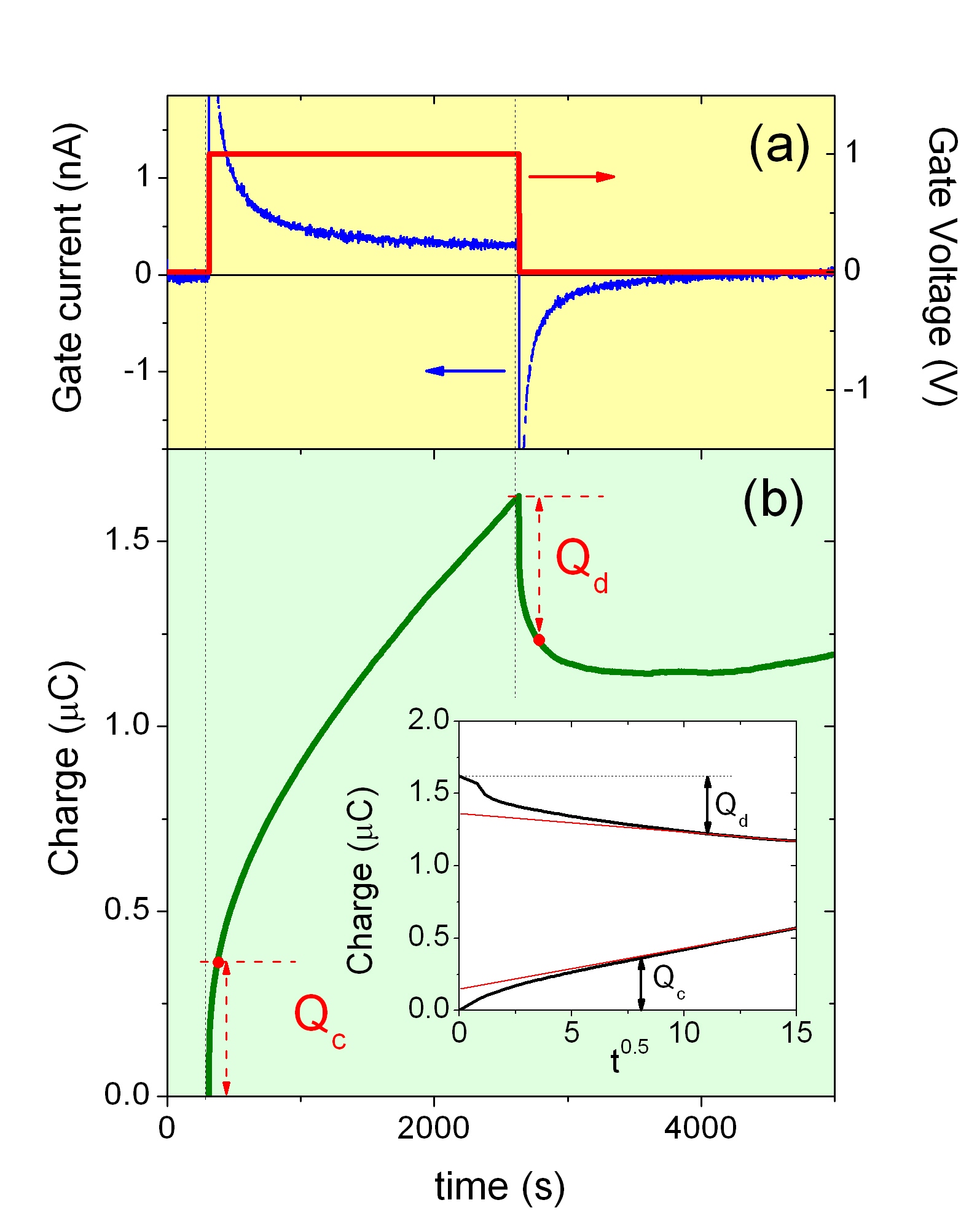}
\end{center}
\vspace{-5mm}\caption{Time dependence of gate voltage and current
(a), and of the charge obtained by integration of the current (b),
when a gate potential of +1 V is applied and removed. The red dots
in (b) indicate the injected charge $Q_c$ and $Q_d$, obtained in the
charge and discharge phases by means of a chronocoulometric
procedure. As shown in the inset, $Q_c = Q(t^*)$, where $t^*$ is the
time at which the $Q(t)$ curve starts to be linear as a function of
$t^{\frac{1}{2}}$. $Q_d$ is determined in a similar way.}
\label{fig:2}
\end{figure}

Figure \ref{fig:2} shows the time dependence of the gate current
$I_G$ (a) and of the total charge $Q(t)=\int_0^t I_G(t') dt'$ (b)
when a gate voltage of 1 V is applied and then removed. The curves
are very similar to the typical ones depicted in \cite{Inzelt10}.
Note that, after the first voltage step, a non-vanishing gate
current continues to flow indefinitely. This current is due to the
flow of charges necessary to maintain the gradient of ion
concentration when tunneling effect through the EDL \cite{Yuan10} or
diffusion of electroreactants \cite{Inzelt10} take place. The shape
of $Q(t)$ shows indeed that two phenomena occur on very different
length scales: a rather fast EDL charging/discharging (that gives
$Q$ an exponential time dependence) and other effects of
electrochemical nature that give a $t^{\frac{1}{2}}$ dependence. In
analogy with the chronocoulometry method, we determined the time
$t^{*}$ at which $Q(t)$ starts to become linear as a function of
$\sqrt{(t)}$, as shown in the inset to Fig.\ref{fig:2}(b), and
assumed that the total charge ``injected'' in the film surface is
$Q(t^*)$. Clearly, two values are obtained, $Q_c$ and $Q_d$, for the
charge and discharge phases. Normally, they coincide within the
experimental uncertainty; this indicates that no adsorption of
reactants or product occurs \cite{Inzelt10}. The injected charge is
finally defined as $Q_i=(Q_c+Q_d)/2$. In the few cases where $Q(t)$
deviates from the aforementioned behavior in one of the two steps
(charge or discharge), $Q_i$ is determined by the other step.

Once $Q_i$ is known, the surface density of injected carriers is
$n_{2D}= Q_i/eS$, where $S$ is the surface of the film covered by
the polymer (gated area) and $e$ is the electronic charge.
%
%
Obviously, the charge distribution on the surface is not exactly 2D.
Within the simplest semiclassical model one can imagine that the
whole charge is injected in a surface layer (whose thickness $\xi$
is of the order of the screening length) and that the film behaves
as the parallel of the perturbed and unperturbed regions. A trivial
free-electron calculation (assuming constant effective electron mass
and relaxation time) of the resistance of the whole film gives
\begin{equation}
\Delta R/R'=\frac{R(V_G)-R_0}{R(V_G)}=-\frac{n_{2D}}{n t}.
\label{eq:DeltaR}
\end{equation}
where $n$ is the unperturbed 3D density of charge carriers. In this
equation, $\Delta R/R'$ does not depend explicitly on $\xi$, but
only on the whole film thickness $t$ and, of course, on $n_{2D}$. A
more sophisticated perturbative self-consistent quantum approach
based on the Lindhard-Hartree theory of the electronic screening
\cite{Omini} and including a proper model of the film conduction
(e.g, accounting for the probability $p$ of electronic specular
reflection at the film surface \cite{Fuchs38}) gives a similar
equation, but with an additional factor that depends in a
complicated way on $t$ and $p$. This term reduces to 1 when $p=0$.
Further details will be given elsewhere \cite{Omini}.

\begin{figure}[t]
\begin{center}
\includegraphics[keepaspectratio, width=\columnwidth]{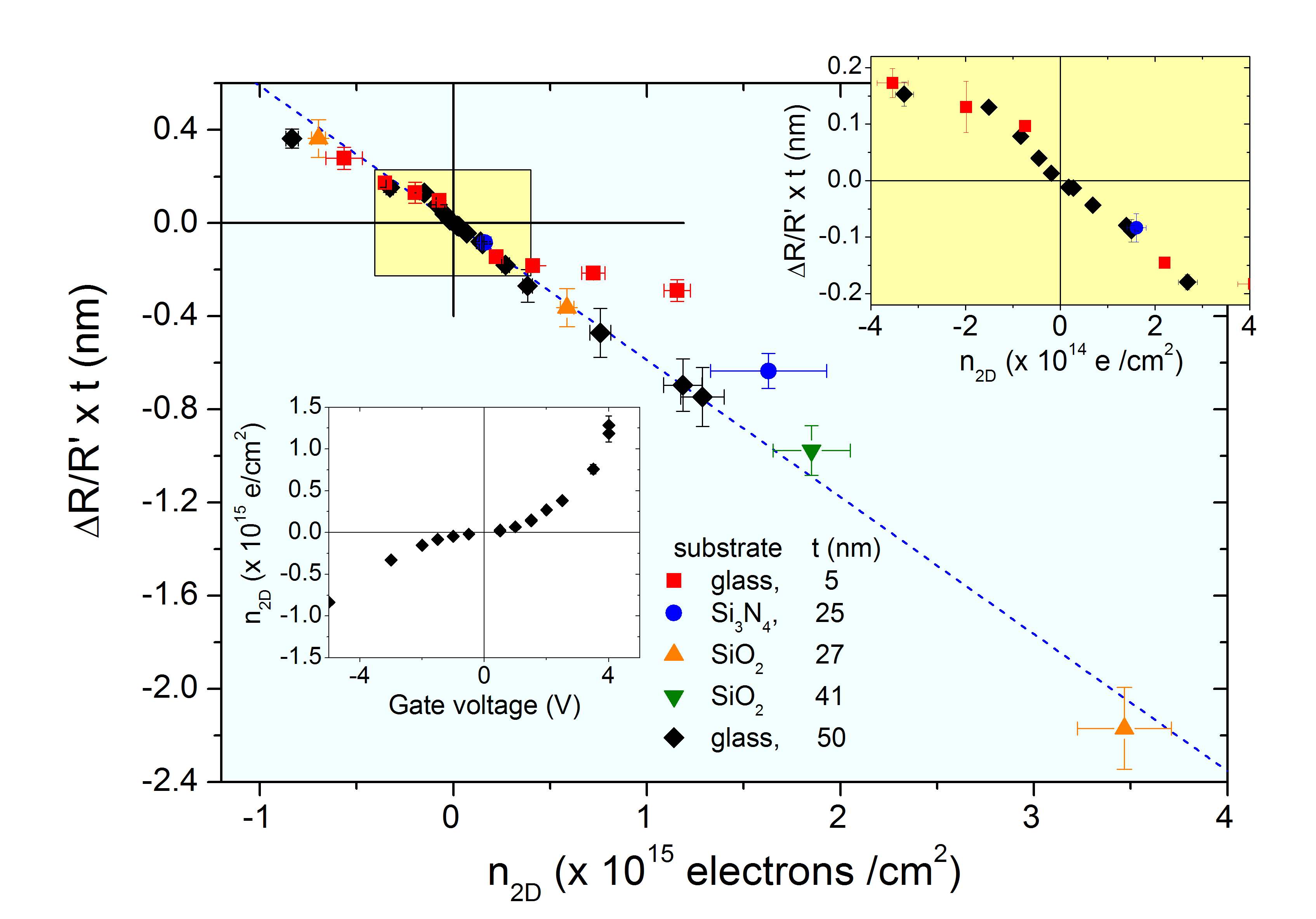}
\end{center}
\vspace{-5mm} \caption{Dependence of $\Delta R/R' t$ on $n_{2D}$
(i.e. number of electrons per cm$^2$) as obtained for various films
with different thickness and on different substrates, indicated in
the legend. The straight dotted line is a guide for the eyes. The
upper inset shows a zoom around the origin of the axes. The lower
inset shows the dependence of $n_{2D}$ on the gate voltage, for the
50-nm thick Au film.} \label{fig:3}
\end{figure}

Figure \ref{fig:3} shows that, for the great majority of the devices
studied here, $\Delta R/R' t$ is a linear function of $n_{2D}$, in
agreement with eq. \ref{eq:DeltaR} \footnote{The thickness of the
thinner film (5 nm) results from a correction that accounts for the
voids in polycrystalline films that reduce the actual film cross
section (see ref.\cite{Rowell03}). In the other cases the correction
was not necessary.}. Vertical and horizontal error bars account for
the difference in the values of $\Delta R/R'$ and $n_{2D}$
determined in the charging and discharging phases -- i.e. on
application and removal of the gate voltage, see fig.\ref{fig:1}
(c,d) and the inset to fig.\ref{fig:2}(b).
%
%
Note that the values of $n_{2D}$ reported so far in literature are
included in the yellow region around the origin of the axes. A
magnification of this region is shown in inset. The values of
$n_{2D}$ obtained with our technique extend instead up to $3.5
\times 10^{15}$ electrons/cm$^2$. The same linear trend is common to
all devices, but some deviations occur in the thinner ones at higher
charge densities. This is not surprising since in these films the
surface scattering plays a major role, and the simple free-electron
model (eq. \ref{eq:DeltaR}) breaks down. A reduction in the absolute
value of $\Delta R/R'$ for a given $n_{2D}$ is indeed predicted by
the aforementioned quantum perturbative model \cite{Omini} when the
probability of electron reflection at the surface \cite{Fuchs38} is
not negligible.

\begin{figure}[t]
\begin{center}
\includegraphics[keepaspectratio, width=\columnwidth]{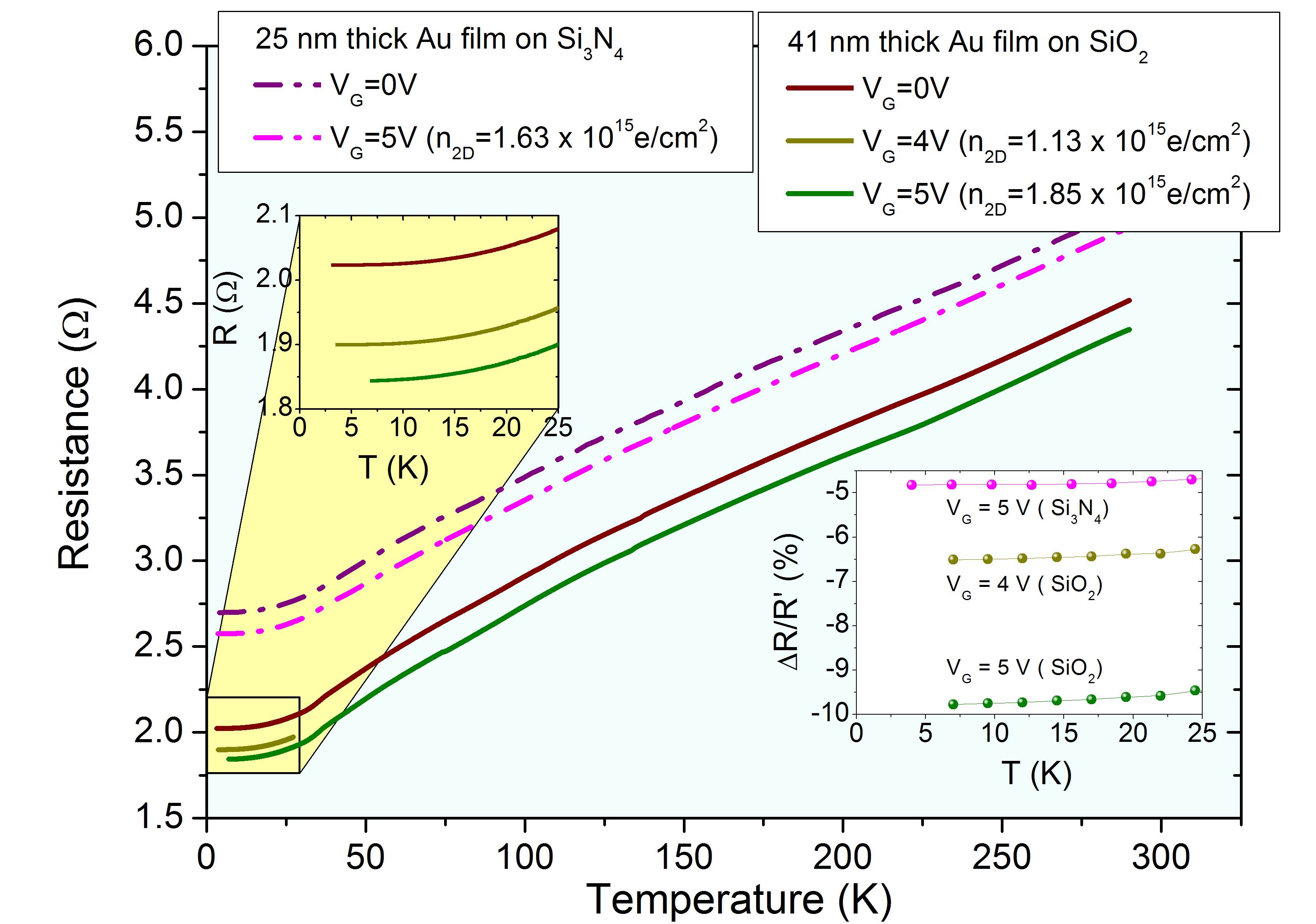}
\end{center}
\vspace{-5mm} \caption{Temperature dependence of the resistance of
two Au films for different values of the gate voltage. The
corresponding values of $n_{2D}$, measured \emph{at room
temperature}, are indicated in the legend. Upper inset: zoom of the
low-temperature region. Lower inset: relative resistance variation
$\Delta R/ R'$ at low temperatures, extracted from the curves in the
main panel. } \label{fig:4}
\end{figure}

In view of the application of this gating technique to more
interesting 2D materials, like graphene and multilayer graphene,
graphane \cite{Giustino10}, MoS$_2$, BN, NbSe$_2$ and so on -- in
particular to see whether some of these materials can develop
superconductivity upon charge doping -- it is important to check
what happens when the device is cooled to cryogenic temperatures.
Because of the glassy transition of the polymer at $T_{glass}\simeq
210$ K, and the consequent ``freezing'' of the EDL charge below that
threshold, the gate voltage must be applied at $T> T_{glass}$ and
kept constant on cooling. As expected, the gate current that
persists after the EDL charge (see fig.\ref{fig:2}(a)) and that is
related to the ionic flow in the polymer electrolyte goes smoothly
to zero on crossing the glassy transition. The cooling speed should
be small enough to avoid cracks in the film or in the contacts due
to the abrupt thermal contraction of the polymer. The resistance of
the film is then measured on slowly heating the FED from the lowest
temperature (here about 3.3 K) to room temperature. Figure
\ref{fig:4} shows the $R(T)$ curves for two Au films on different
substrates, i.e. Si$_3$N$_4$ (dash-dot lines) and SiO$_2$ (solid
lines). The curves at $V_G=0$ and $V_G =5$ V are shown for both
devices; for the latter, an additional curve at $V_G=4$V is
reported, though it extends only up to 28 K because one of the
contact broke down at that temperature. A large offset is observed
within each series, due to the applied field. The inset shows the
low-temperature values of $\Delta R/R'$ extracted from these curves.
At the lowest temperatures, the resistance varies by almost 10\%,
which is a huge quantity for a noble metal. Incidentally,
preliminary measurements on Cu films indicate an even larger effect
(up to 30\%).

\begin{figure} [t]
\begin{center}
\includegraphics[keepaspectratio, width=1\columnwidth]{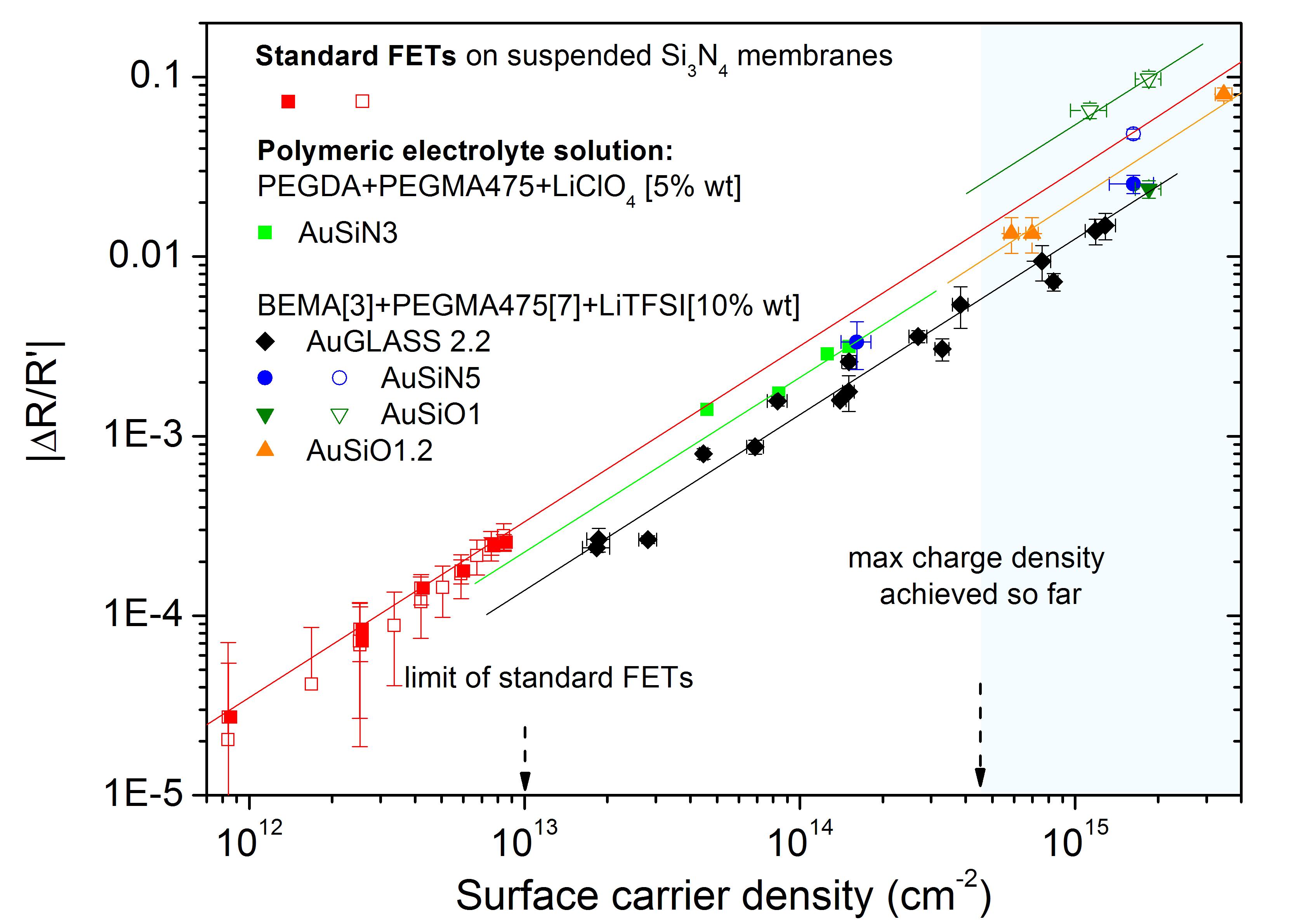}
\end{center}
\vspace{-5mm} \caption{Logarithmic plot of $|\Delta R/R'|$ vs. the
surface density of charge carriers in standard back-gate FETs and in
devices made with two different kinds of PES. Solid (open) symbols
indicate data taken at room temperature (low
temperature).}\label{Fig:Audifftech}
\end{figure}

Finally, fig. \ref{Fig:Audifftech} reports and compares in log-log
scale some results obtained in Au devices of different kinds, i.e.
based on PESs with different compositions, and also in conventional
back-gate field-effect devices made by depositing the Au film and
the electrodes on top of a suspended SiN membrane \cite{Delaude09}
with the Au gate electrode on the other side. The figure clearly
shows
that $|\Delta R/R'|$ is a linear function of $|n_{2D}|$ for all
kinds of devices; the vertical offset of the parallel trend lines is
mainly due to the different thickness of the films.

In conclusion, we have shown that with a suitable polymeric
electrolyte solution it is possible to extend the range of surface
charge densities achieved in field-effect experiments (even at
cryogenic temperatures) to some units in $10^{15}$ charges/cm$^2$.
These values are well in the range where giant modulations of the
conduction properties of some 2D materials, and even the occurrence
of superconductivity (e.g. in graphane \cite{Giustino10}) are
expected. For the time being, we have shown that these huge carrier
injections give rise to large variations in the resistance of Au
thin films, up to about 10\%. The quantity $\Delta R/R'$ for a given
device linearly depends on $n_{2D}$, while \emph{all} the data
follow a universal linear trend if a proper normalization to the
film thickness is used. Some deviations are observed in very thin
films, where the free-electron model is unable to describe the
conduction. These deviations are however compatible with more
sophisticated perturbative quantum models.

\end{document}